\documentclass[twoside]{article}
\usepackage[latin1]{inputenc}
\usepackage{t1enc}
\usepackage{a4}
\usepackage{tabularx}
\usepackage[varg]{txfonts}
\usepackage{graphicx}
\usepackage{natbib}
\usepackage{verbatim}

\begin{document}
\sloppy
\setcounter{figure}{0}
\setcounter{section}{0}
\setcounter{equation}{0}
\setcounter{page}{1}


\begin{center}

{\Large\bf
Multi-dimensional numerical simulations of\\[0.2cm]
type Ia supernova explosions}\\[0.7cm]

F.~K.~R{\"o}pke \\[0.17cm]
Max-Planck-Institut f{\"u}r Astrophysik \\
Karl-Schwarzschild-Str.~1, 85741 Garching, Germany \\
\vspace{0.1cm}

\texttt{fritz@mpa-garching.mpg.de\\
www.mpa-garching.mpg.de/}\verb1~1\texttt{fritz}
\end{center}

\vspace{0.5cm}

\setcounter{figure}{0}
\setcounter{section}{0}
\setcounter{equation}{0}

\def\aj{AJ}%
\def\araa{ARA\&A}%
\def\apj{ApJ}%
\def\apjl{ApJ}%
\def\apjs{ApJS}%
\def\ao{Appl.~Opt.}%
\def\apss{Ap\&SS}%
\def\aap{A\&A}%
\def\aapr{A\&A~Rev.}%
\def\aaps{A\&AS}%
\def\azh{AZh}%
\def\baas{BAAS}%
\def\jrasc{JRASC}%
\def\memras{MmRAS}%
\def\mnras{MNRAS}%
\def\pra{Phys.~Rev.~A}%
\def\prb{Phys.~Rev.~B}%
\def\prc{Phys.~Rev.~C}%
\def\prd{Phys.~Rev.~D}%
\def\pre{Phys.~Rev.~E}%
\def\prl{Phys.~Rev.~Lett.}%
\def\pasp{PASP}%
\def\pasj{PASJ}%
\def\qjras{QJRAS}%
\def\skytel{S\&T}%
\def\solphys{Sol.~Phys.}%
\def\sovast{Soviet~Ast.}%
\def\ssr{Space~Sci.~Rev.}%
\def\zap{ZAp}%
\def\nat{Nature}%
\def\iaucirc{IAU~Circ.}%
\def\aplett{Astrophys.~Lett.}%
\def\apspr{Astrophys.~Space~Phys.~Res.}%
\def\bain{Bull.~Astron.~Inst.~Netherlands}%
\def\fcp{Fund.~Cosmic~Phys.}%
\def\gca{Geochim.~Cosmochim.~Acta}%
\def\grl{Geophys.~Res.~Lett.}%
\def\jcp{J.~Chem.~Phys.}%
\def\jgr{J.~Geophys.~Res.}%
\def\jqsrt{J.~Quant.~Spec.~Radiat.~Transf.}%
\def\memsai{Mem.~Soc.~Astron.~Italiana}%
\def\nphysa{Nucl.~Phys.~A}%
\def\physrep{Phys.~Rep.}%
\def\physscr{Phys.~Scr}%
\def\planss{Planet.~Space~Sci.}%
\def\procspie{Proc.~SPIE}%
\let\astap=\aap
\let\apjlett=\apjl
\let\apjsupp=\apjs
\let\applopt=\ao

\begin{abstract}
\noindent \textit{
The major role type Ia supernovae play in many fields of astrophysics
and in particular in cosmological distance determinations calls for
self-consistent models of these events. Since their mechanism is
believed to crucially depend on phenomena that are inherently
three-dimensional, self-consistent numerical models of type Ia
supernovae must be multi-dimensional. This field has recently seen a
rapid development, which is reviewed in this article. The different
modeling approaches are discussed and as an illustration a
particular explosion model -- the deflagration model -- in a specific
numerical implementation is presented in greater detail. On this exemplary case, the
procedure of validating the model on the basis of comparison with
observations is discussed as well as its application to study
questions arising from type Ia
supernova cosmology.
}
\end{abstract}

\section{Introduction}
The fact that many astrophysical processes are inherently
three-dimensional makes realistic numerical simulations challenging.
Although computational resources increase steadily, complex phenomena
cannot be directly resolved in such simulations in the foreseeable
future. Even exploiting the available computational power and memory,
as most
multi-dimensional astrophysical simulations do, considerable effort is required in
modeling the processes in a way that despite the shortcomings in
resolution the results provide credible physical approximations to the
problems. Moreover, the implementation of efficient numerical
techniques is one of the cornerstones of successful multi-dimensional
modeling of astrophysical phenomena.

Type Ia supernovae (SNe~Ia henceforth) are an excellent example for this
class of astrophysical objects. Numerical simulations of these events
in three dimensions easily reach the
limits of today's computational resources. But as the
underlying physical mechanism is believed to crucially depend on
three-dimensional phenomena, such as turbulence, this is the only
conceivable way towards self-consistent models of type Ia supernova
explosions.
Since \citet{hillebrandt2000a} reviewed type Ia
supernova explosion
models, the field has witnessed a brisk development. Progress in
numerical methods and computational capabilities facilitated
pioneering three-dimensional (3D) simulations, which will be reviewed in
the following.

Self-consistent SN~Ia models are called for in order to achieve a sound
understanding of the mechanism of these astrophysical events.
This is motivated by their significant impact on
many aspects of astrophysics and cosmology. Being one of the main sources
of iron group elements, SNe~Ia contribute to the chemical
evolution of galaxies \citep[e.g.][]{francois2004a}. They affect star
formation and drive shock waves in the interstellar and intergalactic
media.

Most remarkable, however, was the application of SNe~Ia in observational
cosmology, where these objects were employed as distance indicators
\citep[as put forward by][]{branch1992a}. Evidently, SNe~Ia are a valuable tool
to extend the Hubble diagram to large redshifts and to determine the
Hubble constant \citep{hamuy1996b, branch1998a}. At redshifts above 0.5, a significant deviation from
the linear Hubble law was noticed
\citep{riess1998a,perlmutter1999a}. Here, SNe~Ia appear dimmer than
expected in a matter-dominated flat or open Friedmann-Robertson-Walker
Universe. This led to the spectacular
interpretation that the expansion of Universe is currently undergoing
an acceleration. The determination of the driving force of this acceleration
is one of the greatest challenges in contemporary physics. Meanwhile
it is parametrized as ``dark energy'' \citep[see e.g.][]{leibundgut2001a}. Determinations of
cosmological parameters based on anisotropies in the
cosmic microwave background radiation \citep{spergel2003a} and on large-scale galaxy
surveys provided independent confirmation of the
SN~Ia measurements.

Yet the question of the applicability of SNe~Ia as distance indicators
is still not satisfactorily answered. SNe~Ia are remarkably uniform
events by astrophysical standards, but evidently no
standard candles. Only a calibration of the distance measurements
according to empirical correlations between observables provides the
necessary accuracy for the determination of cosmological parameters.
A firm theoretical reasoning of such correlations is, however, still
lacking.

The simplest form of dark energy is a
cosmological constant, but more complicated contributions to the
energy-momentum tensor in the Einstein equations are also
conceivable. How can one determine the nature of the dark energy? A
first step would be to constrain its equation of state. SNe~Ia seem to be
a suitable tool for this task and currently two major campaigns
\citep{astier2005a, sollerman2005a} apply them in distance
determinations of hundreds of supernovae out to redshifts of $z \sim 1$ and
systematic satellite-borne SN~Ia observations are planned.
The large number of observations is necessary to reduce the
statistical errors because putting tight constraints on dark energy
equation of state \citep[see][for recent results]{astier2005a} requires a high accuracy of the distance
determinations.  But still, potential systematical errors
arising from calibration of the distance measurements may obscure the
results. Getting a handle on these is one of the goals of modeling
SNe~Ia.
Obviously, the predictive power of self-consistent multi-dimensional
SN~Ia models is promising for progress in answering the questions
arising from SN~Ia cosmology.

In Sect.~\ref{sect:astro} we set out how the astrophysical scenario
of SNe~Ia is derived from observations. Numerical approaches to modeling
SNe~Ia are discussed in Sect.~\ref{sect:num_ap} with special emphasis
on multi-dimensional models. As an illustrative example, a particular
implementation of the deflagration SN~Ia model is presented in
Sect.~\ref{sect:def}, where also ways to test the validity of SN~Ia
models on the basis of comparison with observations are outlined and
the application to questions arising from the cosmological application
of SNe~Ia is discussed. Conclusions are drawn in Sect.~\ref{sect:concl}.

\section{Observations and astrophysical scenario}
\label{sect:astro}
Observational features of SNe~Ia suggest a specific astrophysical
scenario. The cornerstones of the astrophysical model of SNe~Ia are
set by two fundamental characteristics of these events. Evidently, SNe
Ia belong to the most energetic cosmic explosions, releasing about
$10^{51} \, \mathrm{erg}$ of energy. For a short period of time
they can outshine an entire galaxy consisting of tens of billions of
stars. SNe~Ia spectra are characterized by the lack of
indications for hydrogen and helium which together with a pronounced P
Cygni silicon line at maximum light classifies these objects
\citep{wheeler1990a}. Lines
of intermediate mass elements (such as Si, Ca, Mg and S) and
oxygen are observed in near-maximum light spectra
\citep[e.g.][]{filippenko1997b,filippenko1997a}. With respect to light
curve and spectra observations SNe~Ia form a class of remarkable
homogeneity \citep[e.g.][]{branch1992a}.

Assuming supernovae to originate from single stellar objects, only
their gravitational binding energy, released in a collapse towards a
compact object \citep{zwicky1938a}, or its nuclear energy, released in
explosive
reactions \citep{hoyle1960a}, come into consideration as possible
energy sources. In the
particular case of SNe~Ia no compact object is found in the remnant
excluding the first possibility. The homogeneity of the class of SNe~Ia
and the fact that no hydrogen is found in
their spectra provides a strong hint that the object
undergoing the nuclear explosion may be a white dwarf (WD) star consisting
of carbon and oxygen (C+O).

Lightcurves of SNe~Ia rise over a time scale of several days and
decline over months. It is therefore clear that they cannot be powered
directly by the explosion since the temperatures fall off much too rapidly in the
expansion. This problem was solved by \citet{truran1967a} and
\citet{colgate1969a} who suggested that the $^{56}$Ni produced in large
amounts in the explosive thermonuclear burning provides the energy
source for the optical event by radioactive decay to $^{56}$Co and
$^{56}$Fe.

\subsection{Progenitor evolution and ignition}
\label{sect:progenitor_ignition}

A single WD is an inert object. How can it reach an explosive state?
The only way to introduce the necessary dynamics into the system is to
assume it to be part of a binary system and to gain matter from the
companion. Several models have been proposed for this progenitor evolution.

In the \emph{double degenerate}  scenario \citep{iben1984a,
  webbink1984a}, two C+O WDs merge. The lighter of the two is disrupted and its matter
accreted onto the heavier WD. To become a potential candidate for a
SN~Ia, the merger should eventually reach the Chandrasekhar-mass (i.e.\ of the
maximum mass that is supported against
gravitational collapse by the pressure of the degenerate electrons). Stellar
evolution predicts that systems fulfilling this requirement should
exist and indeed a potential candidate that will merge in less than
one Hubble time has been detected
\citep{napiwotzki2005a}. Mergers of two C+O WDs provide a
natural explanation for
the absence of hydrogen. However,
numerical simulations indicate that the high accretion rate onto the
more massive WD leads to an off-center ignition and the subsequent
burning could convert the material to oxygen, neon, and magnesium. A WD
of this composition, however, tends to undergo a gravitational
collapse rather than a thermonuclear explosion
\citep[e.g.][]{saio1985a,saio1998a}.

In the \emph{single degenerate} scenario
\citep{whelan1973a,nomoto1982a,iben1984a}, the WD accretes matter from
a non-degenerate companion (either a main sequence or an AGB
star).
This idea was recently supported by the detection of the potential
companion of Tycho Brahe's 1572 supernova
\citep{ruiz-lapuente2004a}, which is a solar-type star.
A detailed analysis of this scenario, however, revealed that
the accretion rates admissible here are restricted to a narrow
window. Too low rates evoke to nova eruptions in which the WD looses
more material than accreted before and too high rates would lead to
the formation of an extended He-rich envelope. Moderate accretion rates build up a
degenerate He-shell which could detonate and trigger a detonation of
the carbon-oxygen core. Since this happens before the WD reaches the
Chandrasekhar-mass limit, such a SN~Ia model is termed
\emph{sub-Chandrasekhar} explosion \citep{woosley1994a}. Somewhat
higher accretion rates, however,
can lead to quiet hydrostatic burning of the accreted material
processing it to carbon and oxygen. In this case, the WD may reach the
Chandrasekhar mass (\emph{Chandrasekhar-mass model},
\citealp{hoyle1960a,arnett1969a,hansen1969a}). Potential
candidates for such systems are
Supersoft X-ray Sources \citep[e.g.][]{kahabka1997a}
The limitation the fuel available in the explosion to the Chandrasekhar mass
($\sim$$1.4\, M_\odot$) makes the described scenario particularly
favorable since it provides a natural explanation for the striking
uniformity of SNe~Ia in the gross observational features. On the other
hand, it is afflicted with great uncertainties. Achieving a stable
mass transfer in the progenitor binary system to build up a
Chandrasekhar mass WD is highly non-trivial
\citep[e.g.][]{nomoto1985a} and the observational
evidence for such systems is sparse.

Although the striking homogeneity holds for most SNe~Ia (so-called
Branch-normals, \citealp{branch1993a}), some events differ
significantly. These explode much weaker (like SN~1991bg, see
\citealt{filippenko1992b, leibundgut1993a}) or more
vigorous (like SN~1991T, see \citealt{phillips1992a,
  filippenko1992a, ruiz-lapuente1992a, jeffery1992a, spyromilio1992a})
than the average. At present it is unclear,
whether the entire class of SNe~Ia can be explained by only one
progenitor scenario. The currently favored Chandrasekhar-mass model
may possibly only be able to explain the Branch-normals while sub- and
superluminous events require different progenitor
scenarios. Therefore, even if currently not in the focus of research,
the double-degenerate and the sub-Chandrasekhar mass scenarios may
contribute to the SN~Ia population. In the following, however, we will
focus on the Chandrasekhar-mass model, since it received most
attention in recent theoretical modeling.

\pagebreak

When the WD  approaches
the Chandrasekhar limit,
the density at the center of the WD increases rapidly so
that fusion of carbon ignites. Contrary to the situation in main
sequence stars, the degenerate material of the WD does not allow for
moderation of the burning by expansion. Heat transport is achieved
here by convection giving rise to a stage of
convective carbon burning that lasts for several hundred years. This
phase is terminated by one or more small spatial regions undergoing a
thermonuclear runaway, marking the birth of a thermonuclear flame
and the onset of the explosion.
The convective burning stage and
the conditions at flame ignition are extremely hard to model both
analytically and numerically. Therefore the exact shape and
location of the first flame spark(s) is not yet well
constrained. These, however, are crucial initial
parameters in multi-dimensional explosion models. Only a few studies
addressed the flame ignition process so far. \citet{garcia1995a}
simulated the thermonuclear runaway of a hot bubble floating upward
from the center of the WD. They concluded that off-center and
multi-spot ignitions are possible. A simular result was recently
obtained by \citet{iapichino2005a}.
The convective phase directly preceding the ignition crucially influences
the configuration of the initial flame. Here different studies led to
controversial results. While \citet{woosley2004a} and
\citet{kuhlen2005a} favor off-center, possibly multiple and
asymmetrically distributed flame sparks, \cite{hoeflich2002a} put forward a single
central ignition.

\subsection{Flame propagation and explosion}
\label{sect:flame_explosion}

The goal of SN~Ia explosion models is to follow the propagation of the
thermonuclear flame from its ignition near the center of the WD
outwards and to determine the composition and the distribution of the
burning products in the ejected material.

Theoretically, SN~Ia explosions are governed by the equations of reactive
fluid dynamics, i.e.\ the Navier-Stokes equations extended with an
equation of species balance,
suitable terms for heat conduction, diffusion of species, and source terms
of energy and species in combination with an equation of state
\citep[eg.][]{oran1987a}. These equations allow for solutions of
traveling reaction waves converting unburnt to burnt material. If the
scales under consideration are much larger than the internal
structures of these waves, a simplified description of the system is
possible. Neglecting all phenomena that actually govern the
propagation of the reaction wave, i.e.\ the reaction kinetics and
transport processes, the burning front is modeled as a moving
discontinuity. This picture provides a description of the
hydrodynamical state of the material in the unburnt and burnt regions
and the equations simplify to the reactive Euler equations. From these,
the Rankine-Hugoniot jump conditions for the state variables accross
the burning front follow \citep[see][]{landaulifshitz6eng}
which allow for two
different modes of front propagation. One is the subsonic deflagration in
which the flame is mediated by the thermal conduction of the
degenerate electron gas and the other is a supersonic detonation in
which the burning front is driven by shock waves.
Either one of these modes or a combination of both have been suggested
in different explosion models.

\pagebreak

\begin{itemize}
\item \emph{The prompt detonation model} was first applied in a numerical
  simulation by \citet{arnett1969a}. A spherically symmetric detonation
  wave was initiated near the center of the WD and propagated
  outward. Criteria for the spontaneous formation of a detonation wave
  have been investigated by \citet{blinnikov1986a} and
  \citet{woosley1990a}. They reach the conclusion, that a detonation
  can initiate only under certain prerequisites. A prompt detonation as explosion model
  produces enough energy for a SN~Ia event. However, ahead of
  a supersonic detonation wave the fuel cannot expand and is therefore
  incinerated at the high densities of an equilibrium white
  dwarf. This results in the almost complete conversion of the
  material to nickel-peaked nuclear statistical equilibrium
  \citep{arnett1969a,arnett1971a}, which is in
  conflict with the intermediate mass elements observed in SN~Ia
  spectra. These nucleosynthetic problems rule out a pure
  detonation scenario as a standard model for SN~Ia explosions.

\item \emph{The deflagration model}
  \citep{nomoto1976a} assumes the flame
  propagating in the subsonic deflagration mode. The laminar burning speed of the
  deflagration flame is determined by microphysical transport
  processes. For conditions of carbon burning in C+O WDs it is highly
  subsonic \citep{timmes1992a} and therefore the flame propagates
  far too slowly to explain SN~Ia explosions. The expansion of the star
  will quench burning before the WD gets unbound. On the other hand, this
  model can cure the problem of nucleosynthesis, since rarefaction
  waves travel ahead of the flame with sound speed and lower the
  fuel density prior to burning. Thus the material can partly be processed into
  intermediate mass elements.

  The deflagration model undergoes a significant improvement when multidimensional
  effects are taken into account.
  The propagation of the deflagration front is
  subject to several instabilities \citep[e.g.][]{niemeyer1997b}. Of purely hydrodynamical origin is
  the Landau-Darrieus instability \citep{landau1944a,darrieus1938a}
  which in the nonlinear stage is stabilized in a cellular pattern
  \citep{zeldovich1966a} thus enlarging the flame surface area and
  enhancing the net burning rate. The major effect accelerating the
  flame, however, is due to the buoyancy unstable
  flame propagation from the center of the star outwards. It leaves
  behind light and hot ashes below
  the dense fuel -- a density statification inverse to the
  gravitational acceleration. In its non-linear stage, the
  Rayleigh-Taylor instability leads to the formation of
  mushroom-shaped burning bubbles raising into the fuel. The Reynolds
  number typical for this situation is as high as $10^{14}$. Clearly,
  shear (Kelvin-Helmholtz) instabilities at the interfaces of these
  bubbles will generate turbulent eddies which then decay to smaller
  scales forming a turbulent energy cascade. The flame will interact
  with these eddies down to the Gibson-scale at which the turbulent
  velocity fluctuations become comparable to the laminar flame
  speed. Below the Gibson scale, the flame burns faster through
  turbulent eddies than they can deform it, and the flame propagation
  is thus unaffected by turbulence there. This interaction
  corrugates the flame again increasing its surface
  and consequently accelerating the effective propagation speed.

\pagebreak

\item \emph{The delayed detonation (DD) model} conjoins the
  advantages of the deflagration and the detonation models. It was put
  forward by \citet{ivanova1974a}, \citet{khokhlov1991a}, and
  \citet{woosley1994a}. Burning
  starts out in the slow deflagration mode
  pre-expanding the star. Motivated by transitions from the
  deflagration to the supersonic detonation mode
  (deflagration-to-detonation transition, DDT) observed in
  terrestrial turbulent combustion processes, such a phenomenon is
  assumed to occur in SNe~Ia. Its physical mechanism, however, remains unclear
  and therefore it enters the model as a free
  parameter. Usually, in one-dimensional (1D) simulations the DDT is
  artificially initiated
  once the  flame reaches fuel of a certain transition
  density $\rho_\mathrm{tr}$. The assumed detonation then
  burns the star until the flame is quenched by the expansion.
  This detonation is an easy way to explain the energy release
  necessary for a SN~Ia explosion. The
  important notion in this model is that a detonation in low density
  fuel (pre-expanded in the deflagration stage)
  can lead to only partial burning and is therefore capable
  of generating intermediate mass elements. Another possible advantage of
  this model derives from a problem of current 3D
  implementations of the pure deflagration scenario. The Rayleigh-Taylor
  bubbles being the origin of the turbulent flame acceleration cause
  considerable amounts of
  unburnt C+O matter to remain in ``fingers'' near the center of the
  star, which are possibly in conflict with spectral SN~Ia
  observations. A detonation wave initiated at later stages of the
  evolution could be capable to burning out
  those fingers and to process the previously unburnt material. The best agreement with
  observations was achieved for
  $\rho_\mathrm{tr} \sim 1 \times 10^7 \, \mathrm{g} \,\mathrm{cm}^{-3}$
  \citep{hoeflich1996a,iwamoto1999a}.
  The main disadvantage of the DD model is that this transition density
  remains an arbitrary parameter unless the mechanism of a possible
  DDT is physically determined \citep[if there exists one at all,
  cf.][]{niemeyer1999a}.

\item \emph{The pulsational delayed detonation (PDD) model}
  \citep{arnett1994a, arnett1994b} is
  similar to the DD model in the sense that it combines an initial
  deflagration with a later detonation. The flame is assumed to
  propagate in the initial deflagration phase with its laminar burning
  speed and pre-expands the star. Due to the slow flame velocity, the
  burning front stalls and fails to unbind the star. The WD then
  re-contracts giving the interface between burnt and unburnt material
  enough time to mix and to become nearly isothermal. Compressional
  heating finally triggers a detonation at densities that are lower
  than that prior to the first expansion phase. \citet{hoeflich1996a}
  employ this scenario in a phenomenological 1D model and conclude
  that it may account for sub-luminous SN~Ia. However, the assumption
  that the flame propagates with the pure laminar burning velocity in
  the deflagration phase seems unrealistic, because of the flame
  instabilities and the resulting turbulent flame acceleration. Recent
  multidimensional deflagration models
  \citep{reinecke2002d,gamezo2003a} demonstrated that taking into
  account these effects, the star is likely to get unbound instead of
  recontracting.
\end{itemize}

\section{Numerical models}
\label{sect:num_ap}

\subsection{Relevant scales}
\label{sect:scales}

The numerical implementation of SN~Ia models accounting for the full
exploding WD star in multiple dimensions is
significantly complicated by the wide range of relevant length scales
involved in the problem. From the radius of the WD star
($\sim$$2000\,\mathrm{km}$ at the onset of the explosion and expanding
in the process) it reaches down to the flame width which is well below one
centimeter. In the deflagration model the relevance of turbulent
effects amplifies the scale problem since the turbulent
cascade extends to the even much smaller Kolmogorov scale where
the turbulent energy is dissipated into heat. Here, the flame
interaction with the turbulent cascade down to the Gibson scale
must additionally be taken into account.
Current 3D simulations capturing the entire star reach
resolutions
around one kilometer while the Gibson scale is of the order of
$10^4\,\mathrm{cm}$ at the beginning of the explosion and decreases
steadily.

For large-scale multi-dimensional SN~Ia simulations this has three consequences.
\begin{enumerate}
\item \label{enum2} The internal flame structure cannot be
  resolved. Thus, an effective flame model has to be applied and
  complementary small scale simulations are required.
\item \label{enum1} It is not possible to fully resolve the interaction of the flame
  with turbulence. Therefore modeling of the effects on unresolved
  scales is necessary.
\item Assumptions about the flame properties at unresolved scales
  (e.g.\ stability below the Gibson scale) have to be validated in
  separate small-scale simulations.
\end{enumerate}

\subsection{Modeling approaches}

Numerical models of SN~Ia explosions have to face three major
challenges. Apart from the vast range of relevant length scales they need
to take into account inherently three-dimensional physical phenomena
and to solve the kinetics of nuclear burning. To meet all these
requirements in a
single simulation will be impossible in the foreseeable
future. Therefore the problem has been tackled in different
approaches.

The first path towards SN~Ia explosion modeling is to
restrict the simulations to only one spatial dimension. Here, in
principle a resolution of the relevant scales is achievable and a
detailed description of the nuclear reactions is feasible.
However, crucial three-dimensional
physical mechanisms are not explicitly taken into account and need to
be parametrized. Although such models were shown to yield good
fits to observations and pioneered the insight into fundamentals of
the explosion process, they lack consistency and are of little
predictive power.

In multi-dimensional simulations, contrariwise, the computational
costs of modeling the explosion hydrodynamics is prohibitive to
directly resolve all relevant scales as well as details of the nuclear
processes. While the latter may be improved in the forthcoming years,
and is meanwhile separated from the actual explosion simulations
still maintaining a reasonable accuracy (see below), even a drastic
increase in computational capabilities will not allow for a resolution
of all scales in multi-dimensional simulations.

A third approach is to study specific effects on a limited range of
spatial scales, to validate assumptions and improve modeling
techniques of the large-scale SN~Ia simulations.

\subsection{One-dimensional simulations}

Although we focus on multi-dimensional models of SN~Ia explosions, we
will first give a brief overview of 1D spherically
symmetric simulations. Since
(at least in the initial deflagration stage) the propagation velocity
of the thermonuclear flame is determined by multi-dimensional effects,
such as instabilities and
interaction with turbulent velocity fluctuations, a physically undetermined
speed is ascribed to the spherical burning front. This free parameter
can be used to fit the observations and therefore give a hint to the
expected average temporal evolution of the flame in multi-dimensional
simulations. The best known of these models, W7, was presented by
\citet{nomoto1984a} and \citet{thielemann1986a} and has been employed
in several studies since
then \citep[e.g.][]{iwamoto1999a, brachwitz2000a}. Here the flame propagation speed
was modeled with mixing length theory and
started slowly to pre-expand the
star. Later it strongly accelerated and consumed large
fractions of the star converting the material to iron-group elements
(mainly $^{56}$Ni) and intermediate mass elements.
Different ways to parameterize the flame propagation velocity were
proposed by \citet{nomoto1976a}, \citet{woosley1990a},
\citet{hoeflich1996a}, and
\citet{niemeyer1997b}. All these models conclude that a flame
acceleration to one third of the sound speed is necessary for
consistency of the results with observations. The problem of
overproduction of neutron-rich iron-group material noted in these
models may be significantly reduced when taking into account revised
electron capture rates \citep{brachwitz2000a}.

When applying 1D models to simulate delayed-detonation
scenarios, the condition for the DDT comes into play as a second
undetermined parameter. Such models are capable of providing excellent fits to
lightcurves and spectra and seem to yield reasonable compositions of
the nucleosynthesis products \citep{woosley1990a,
  hoeflich1996a,khokhlov1991b,iwamoto1999a}.

Although lacking consistency and therefore predictive power,
1D models have been employed to address questions arising
from SN~Ia cosmology, such as the origin of the diversity in these
objects \citep{bravo1993a, bravo1996a,hoeflich1996a,
hoeflich1998a, umeda1999b, iwamoto1999a, dominguez2000a,
dominguez2000b, dominguez2001a}.

\subsection{Multi-dimensional simulations}

Initiated by \citet{mueller1982a,mueller1986a}, multi-dimensional models were
applied to fix the turbulent flame propagation velocity undetermined
in spherically symmetric simulations. Given the wide range of scales
on which the
flame is affected by turbulence, this is an ambitious project,
additionally challenged by the lack of resolution of the thermonuclear
flame structure. For both problems, different approaches have been
taken. All these were guided by the theory of turbulent
combustion in terrestrial flames \citep[see][]{peters2000a}.

\subsubsection{Hydrodynamics}

The hydrodynamical equations are discretized either in a Lagrangean or
Eulerian approach. The Smooth Particle Hydrodynamics (SPH) technique used by
\citet{garcia1998a} falls into the first category, while an implicit
scheme on an Eulerian grid is applied by \citet{livne1993a}. Most explicit
Eulerian approaches were based on versions of the \textsc{Prometheus}
implementation \citep{fryxell1989a} of the Piecewise Parabolic Method
(PPM) proposed by \citet{colella1984a}. These include the simulations
of
\citet{arnett1994a,arnett1994b}, \citet{khokhlov1995a},
\citet{niemeyer1995b,niemeyer1996a}, \citet{reinecke1999b},
\citet{hillebrandt2000b}, \citet{reinecke2002d},
\citet{gamezo2003a,gamezo2004a}, \citet{roepke2004c}, \citet{calder2004a},
\citet{plewa2004a},
\citet{roepke2005c}, \citet{roepke2005b,roepke2005a}, \citet{roepke2005d,roepke2005e}
and \citet{gamezo2005a}.
An alternative approach was taken by \citet{bell2004c}, who modified
the hydrodynamical equations to account for low-Mach number flows
only. This provides an efficient scheme to numerically simulate
subsonic deflagrations.

The facts that the WD star is expanding and that the
flame is not necessary domain-filling in all stages of the explosion
is taken into account in the grid-based schemes in various approaches. While
\citet{reinecke1999b, reinecke2002d} use a static computational grid
with a fine-resolved central part and a coarse outer grid to account
for the expansion, \citet{roepke2005c} implemented a uniform moving
grid that co-expands with the exploding WD. Since on average the
explosion process is more or less spherical, one can gain resolution
in the part of the star occupied by the flame with nested moving grids
\citep{roepke2005e}. A different technique providing resolution where
needed is adaptive mesh refinement as applied in
the simulations by \citet{gamezo2003a} and \citet{calder2004a}.

\subsubsection{Flame model}
In earlier works \citep{mueller1986a,livne1993a} the flame was
advanced according to discrete boolean criteria. A computational
cell was burned if certain conditions in neighboring cells were
fulfilled. This strategy is clearly dependent on the grid geometry
and therefore in recent simulations different approaches have been taken.
Two major strategies to tackle the problem of the unresolved
internal flame structure can be distinguished.
\citet{khokhlov1993a,khokhlov1994a,khokhlov1995a} adapted a flame
capturing technique that mimics flame propagation by an artificial
diffusion mechanism which broadens the internal flame structure to a
certain number
of computational grid cells. This method was applied in the SN~Ia
explosion simulations by
\citet{gamezo2003a,gamezo2004a,gamezo2005a}, \citet{calder2004a}, and
\citet{plewa2004a}. A completely different approach was taken by
\citet{reinecke1999a}, who treat the flame as a sharp discontinuity
separating the fuel from the ashes. It is numerically represented
applying the level-set technique \citep{osher1988a,sethian1996a}. Here
the flame front is associated with the zero level set of a scalar
function $G$ representing the distance from the interface.
A model for flame propagation based on this technique was developed by
\citet{smiljanovski1997a} and a modified version for thermonuclear
flames in SN~Ia explosion was presented by \citet{reinecke1999a} \citep[for
details of the implementation see also][]{hillebrandt2005a}. This
scheme was applied in the simulations by \citet{hillebrandt2000b},
\citet{reinecke2002c,reinecke2002b,reinecke2002d},
\citet{roepke2004c,roepke2005a,roepke2005b}, \citet{roepke2005c},
 \citet{schmidt2005c}, \citet{schmidt2005a}, and \citet{ roepke2005d, roepke2005e}.

\subsubsection{Turbulent combustion model}
\label{sect:turb_comb}

As outlined above, SN~Ia explosions are believed to at least start out
subsonically in the deflagration mode of flame propagation. Thus
the flame is subject to the interaction with turbulence generated by
instabilities. This mechanism turns out to be crucial to accelerate
the flame propagation and must therefore be taken into account in any
valid SN~Ia model.

The theory of turbulent combustion has been extensively developed for
terrestrial combustion phenomena \citep[see][]{peters2000a}.
The nuclear reaction kinetics in thermonuclear combustion waves is
much simpler than the chemical reactions in terrestrial flames and
due to the high thermal conductivity of the degenerate electron gas in
WD matter some of the characteristic nondimensional numbers
differ. Nonetheless, the turbulent combustion process in the
deflagration mode in SNe~Ia bears striking similarity to \emph {premixed
  turbulent flames} (i.e.\ turbulent flames
in mixtures of fuel and oxidizer) in terrestrial combustion processes.

The wide range of scales involved in turbulent combustion phenomena
renders direct simulations impossible for most situations. Therefore
only parts of the interaction range of the flame with turbulence and
the resulting surface enlargement of the flame can be
resolved. This is usually compensated by attributing an effective
\emph{turbulent flame speed} $s_\mathrm{t}$ to the unresolved flame
front, which must be determined by theoretical considerations.
One of the cornerstones of the theoretical description of turbulent
combustion is the notion of different regimes of flame/turbulence
interaction \citep{niemeyer1997d}. These regimes are distinguished by
the ability of
turbulent eddies to penetrate the internal flame structure. Since the
Gibson scale is much larger than the flame width for most parts of the
SN~Ia explosion, this will not be the case here and accordingly the
combustion falls into the regime of \emph{wrinkled and corrugated
  flamelets}. Here, the full flame structure is corrugated by the
interaction with turbulence and the resulting surface enlargement
accelerates its propagation. As first noted by
\citet{damkoehler1940a}, the flame propagation in this regime
completely decouples from the microphysics of the burning for
sufficiently strong turbulence. It is
entirely determined by the turbulent velocity fluctuations, that is,
$s_\mathrm{t}$ is proportional to the turbulent velocity
fluctuations.

One of the challenges of deflagration models of SN~Ia
explosions is thus to determine these velocity fluctuations
correctly. Since the
resolution in multi-dimensional simulations is insufficient to resolve
the phenomena directly, modeling approaches have to be taken. In the
simplest models, a minimal propagation velocity of the flame is
artificially imposed, which is not physically determined in the model
\citep{garcia2005a}. Contrary to this, \citet{gamezo2003a},
and \citet{calder2004a} assume the flame to be driven by
buoyancy-induced instabilities on unresolved scales giving rise to
a turbulent flame speed of
$s_\mathrm{t} = 0.5 \sqrt{A g l}$, where $g$ and $l$ denote the
gravitational acceleration and the computational grid cell size,
respectively. $A$ is the Atwood number derived from the density contrast
over the flame front. A more sophisticated approach was proposed by
\citet{niemeyer1995a}.  Guided by the technique of \emph{Large Eddy
  Simulations} they implemented a turbulent subgrid-scale model
\citep{clement1993a} in SN~Ia explosion simulations. This model
determines the turbulence energy on unresolved scales based on
conservation laws. The hyperbolic nature of the equations of
hydrodynamics, however, does not allow for a closed system of
equations here so that closure assumptions have to be invoked. A
physically better motivated approach based on localized closures was
recently proposed by \citet{schmidt2005b,schmidt2005c}.

In the very late stages of the SN~Ia explosion the fuel
density drops  due to
expansion of the WD to values where the flame width becomes broader than the
Gibson length. Then, turbulence penetrates the internal structure of
the flame and it enters the regime of \emph{distributed
  burning} \citep{niemeyer1997d}. Here, different scaling laws for the
turbulent flame speed apply \citep{damkoehler1940a} and a preliminary
test of the effects on SN~Ia models was given by \citet{roepke2005a}.

\subsubsection{Nuclear reactions}
\label{sect:nuclear}

The computational expenses for a full nuclear reaction network are
prohibitive to run it concurrently with the explosion hydrodynamics
simulations in present multi-dimensional models. Since here the only
dynamically relevant parameter is the energy release in the reactions, 
it is sufficient to apply a simplified description. Usually, only a
few nuclei representative for the fuel mixture, the iron group elements and the
intermediate mass elements in the ashes and effective
reactions between them are accounted for \citep[e.g.][]{reinecke2002b}.

However, in order to derive observables (such as lightcurves and spectra) from
the explosion models, the exact chemical composition of the
ejecta needs to be known. This can be achieved by advecting a number
of tracer particles in the explosion hydrodynamics simulations which
record the evolution of temperature and density. This adds a Lagrangean
component to the Eulerian code. Based on the data gained from the
tracer particles it is then possible to \emph{a posteriori}
reconstruct the detailed nuclear reactions in the burnt material with
extended nuclear reaction networks. This technique was implemented for
SNe~Ia by \citet{travaglio2004a} and also applied in the study by
\citet{roepke2005d}.

\subsubsection{Results of multi-dimensional models}

While earlier two-dimensional (2D) deflagration models of SN~Ia failed to
explode \citep[e.g.][]{khokhlov1995a} others succeeded to (weakly)
unbind the WD \citet{niemeyer1996a}. Nowadays, it seems that a
consensus has been reached on the general capability of the 3D version
of this model to
give rise to robust explosions
\citep{reinecke2002d,gamezo2003a,roepke2005b}. Although the energetics
(asymptotic kinetic energy up to $7 \times 10^{50}\, \mathrm{erg}$) and
the production of $^{56}$Ni ($\sim$$0.4 \, M_\odot$) as the main global characteristcs fall
into the range of observational expectations \citep{contardo2000a,
  stritzinger2005a}, they are still on the weaker side of ``normal''
SNe~Ia.

Since the initial flame configuration in the
explosion is undetermined yet (see
Sect.~\ref{sect:progenitor_ignition}), several studies addressed its
effect on the outcome of the simulations. Off-center ignitions were
analyzed based on 2D-simulations by \citet{niemeyer1996a} and
3D-full star simulations by \citet{calder2004a}. The latter study
indicated that a single perfectly spherical off-center initial flame
quickly emerges to the surface of the WD and fails to burn sufficient
material to explode it. \citet{plewa2004a} suggested that the material
breaking through the surface of the still gravitationally bound WD may
collimate on the opposite side of it, evoking a
detonation there. Contraty to that, \citet{roepke2005b} found
that a more structured off-center initial flame (motivated by pre-ignition
convection, cf.\ Sect.~\ref{sect:progenitor_ignition}) stays more or
less in place and gives rise to a viable explosion. The effects of the
number of ignition spots have been addressed in several 2D studies
\citep{niemeyer1996a,reinecke1999b,livne2005a} giving a wide range of
results dependent on the number and distribution of the initial flame
kernels and in 3D simulations \citep{reinecke2002d,roepke2005b,roepke2005e,
  garcia2005a}.
These simulations indicate that multi-spot ignition scenarios can lead
to more vigorous explosions than centrally ignited setups and put the
deflagration scenario in
better agreement with observations.
Interestingly, the 3D simulations, although also
exhibiting a depencence on the number of ignition spots, gave much
more robust results than the 2D studies \citep{roepke2005e}. Similar
conclusions were drawn on the basis of a stochastic ignition model in
which a spatial and temporal distribution of the flame ignition spots
was assumed \citep{schmidt2005a}.

A successful explosion,
however, is not the only requirement for a valid SN~Ia model. Observables
derived from such models must match the observations of
SNe~Ia. Moreover, the model must allow for a certain variation of the
results in order to
reproduce the observed range of diversity in the characteristics.
No general consensus has been reached so far regarding these
questions. Besides the low explosion energies and $^{56}$Ni production
the problems of most deflagration simulations include unburnt material
left behind at low velocities in the ejecta due to downdrafts in the
buoyancy-induced large-scale flame pattern and low production of
intermediate mass elements. Recently, a series of 3D
simulaions was analyzed to determine wheter these
shortcomings are generic to the deflagration model or caused by the
simplicity of some setups. It seems that the model is generally capable of
reproducing main observational features, but some of the problems
persist (see Sect.~\ref{sect:def} for details).

\citet{gamezo2004a} claim that
a delayed detonation is necessary to achieve agreement with
observations. In their 3D model, a detonation front is artificially
initiated after a deflagration stage at a pre-selected time and
location. As expected, it travels
through the WD burning most of the materinal in the inner parts. A 2D
implementation of the delayed detonation model was recently presented
by \citet{golombek2005a}. Here, the DDT is assumed to occur when the
deflagration flame enters the distributed burning regime which
appatently happens first in the outer parts of the flame front. This
rises the question of whether the detonation (if not allowed to travel
through ashes) can catch up with the expanding material on the
opposite side of the WD.
2D simulations by \citet{steinmetz1992a} studied pure prompt
detonation in rapidly rotating WDs. Although here regions may exist
with densities allowing for the production of intermediate mass
elements, the authors conclude, that the ratio of iron group elements
to intermediate mass elements arising from such models is inconsistent with normal
SNe~Ia.

To settle the question of the explosion mechanism,
detailed analysis of the differnt models and comparison with
observations on the basis of synthetic light curves and spectra are
required.

\subsection{Complementary small-scale simulations}
Complementary studies of flame propagation on small scales focus on
effects in only a narrow window in scale space. These therefore reach a
much higher resolution and can model the processes in a more realistic
way than  SN~Ia
simulations on scales of the WD star. Consequently, they
give insight into mechanisms that are unresolved there either
testing the validity of assumptions or providing input
data to these models.

On the basis of resolved 1D flame simulations,
\citet{timmes1992a} determined the laminar speed of deflagration flames in
SNe~Ia and \citet{sharpe1999a} analyzed the propagation velocity of
detonations. Both serve as input in multi-dimensional simulations
\citep{reinecke2002d,golombek2005a}.

Other studies concerned the validity of subgrid-scale models of
the interaction of the flame with turbulence \citep{schmidt2005e}.
This interaction was also simulated by \citet{niemeyer1999b}.
The deflagration flame propagation on small scales in SNe~Ia subject to
the Rayleigh-Taylor instability was tested in 2D-simulations by
\citet{khokhlov1993a} and \citet{bell2004b} and in 3D-setups by
\citet{khokhlov1995a} and \citet{zingale2005a}. The effects of the Landau-Darrieus
instability on the thermonuclear flame have been analyzed in
2D-simulations by
\citet{niemeyer1995a}, \citet{blinnikov1996a}, \citet{roepke2003a}, \citet{roepke2004a,
  roepke2004b}, and \citet{bell2004a}. \citet{dursi2003a} studied the
response of thermonuclear flame to curvature and stretch.
Most of these simulations confirm the assumptions made in large-scale
SN~Ia models.

The question whether a transition of the turbulent burning regime from
flamelet to distributed burning would cause a
deflagration-to-detonation transition was addressed by
\citet{lisewski2000a,lisewski2000b}, who conclude that such an event is
very unlikely.

\section{Example: A deflagration SN~Ia model}
\label{sect:def}

To illustrate the numerical modeling of SN~Ia explosions in three spatial
dimensions, a deflagration model in a specific implementation shall be
discussed in this section. It starts out with a cold isothermal ($T =
5 \times 10^5 \, \mathrm{K}$) Chandrasekhar-mass WD. Unless otherwise
stated, its composition was assumed to consist of equal parts of
carbon and oxygen.

\subsection{Numerical techniques}

The numerical techniques forming the foundation of the implementation of the
deflagration SN~Ia model presented here are described by
\citet{reinecke1999a}, \citet{roepke2005c}, \citet{schmidt2005b}, and
\citet{niemeyer1995b}. The fundamental approach is that of \emph{Large Eddy
Simulations}, where hydrodynamics on the resolved scales is modeled in a \emph{finite volume
approach} based on the \textsc{Prometheus} implementation \citep{fryxell1989a}.

The equation of state describing WD material contains contributions
from an arbitrarily degerneate and relativistic electron gas, an ideal
gas of nuclei, radiation, and possibly electron-positron pair creation and
annihilation.

As a consequence of \ref{enum1} of the enumeration in Sect.~\ref{sect:scales}, a
\emph{subgrid-scale model} is applied to account for turbulent effects on
unresolved scales. Some older simulations follow the implementation
suggested by \citet{niemeyer1995b}, while one recent highly resolved run applied the
updated modeling approach of \citet{schmidt2005b}.

According to \ref{enum2} of the enumeration, the flame representation
is achieved in a modeling approach. Seen from the scales of the WD
star, the flame
appears as a sharp discontinuity separating the fuel from the
ashes. A suitable numerical method to follow the evolution of such an
interface is the \emph{level set technique} introduced to SN~Ia
modeling by \citet{reinecke1999b}.
In this implementation, the flame propagation speed needs to be
prescribed. This quantity, however, is not arbitrary in the presented
modeling framework.
For turbulent combustion in the flamelet regime (cf.\
Sect.~\ref{sect:turb_comb}), which
applies to the burning in major parts of SN~Ia explosions, it is given
by the turbulent velocity fluctuations. These are determined by the
subgrid-scale turbulence model.

Nuclear reactions are implemented in the simplified approach (cf.\
Sect.~\ref{sect:nuclear}) proposed
by \citet{reinecke2002b}. The progenitor material is composed of $^{12}$C and
$^{16}$O. At high fuel densities nuclear burning
terminates in nuclear statistical equilibrium represented by a mixture
of $^{56}$Ni and $\alpha$-particles. Once the fuel density drops below
$5.25 \times 10^7 \, \mathrm{g} \, \mathrm{cm}^{-3}$, burning will become
incomplete and intermediate mass elements (represented by  $^{24}$Mg)
are produced. The respective difference in nuclear binding energy is
released which provides sufficient accuracy to model the dynamics of
the explosion.
The chemical composition of the ejecta is derived in a postprocessing
step \citep{travaglio2004a}.

\subsection{Results of Simulations}
\label{sect:results}

\begin{figure}[t!]
\centerline{
\includegraphics[width = 0.75 \linewidth]
  {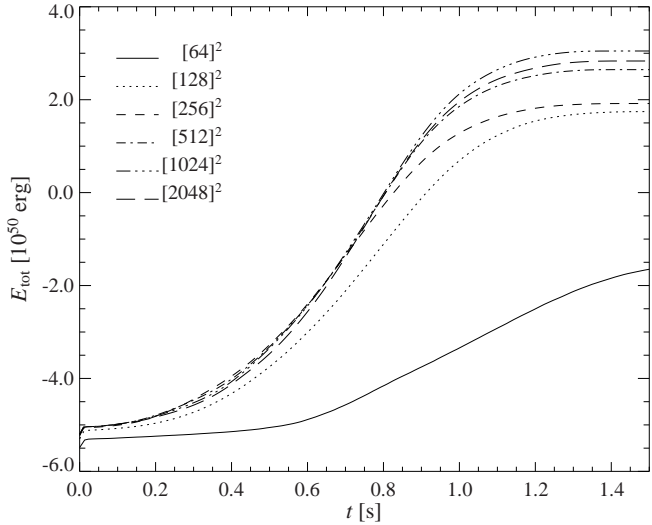}
}
\caption{Total energy in 2D-simulations with co-expanding grid for
  different numbers of computational grid cells. \label{fig:conv}}
\end{figure}

Several simulations based on the implementation described above, both
in two and three spatial dimensions, have
been presented by \citet{reinecke1999b,reinecke2002b,reinecke2002d}. In the
2D simulations, numerical convergence in the global
quantities was demonstrated. For the implementation on a co-expanding computational
grid, a similar result was found by \cite{roepke2005c} (see
Fig.~\ref{fig:conv}). The numerical convergence naturally arises from
the interplay of the resolved flame front representation with the
turbulent subgrid-scale model. Ideally, a lack of resolution of
large-scale features in the flame front representation should be
compensated by an increased turbulent flame propagation velocity
determined from the subgrid-scale approach. Of course, a certain
threshold of resolution will need to be exceeded to reach this regime
in the numerical implementation.

\begin{figure}[t!]
\centerline{
\includegraphics[width = 0.9 \linewidth]
  {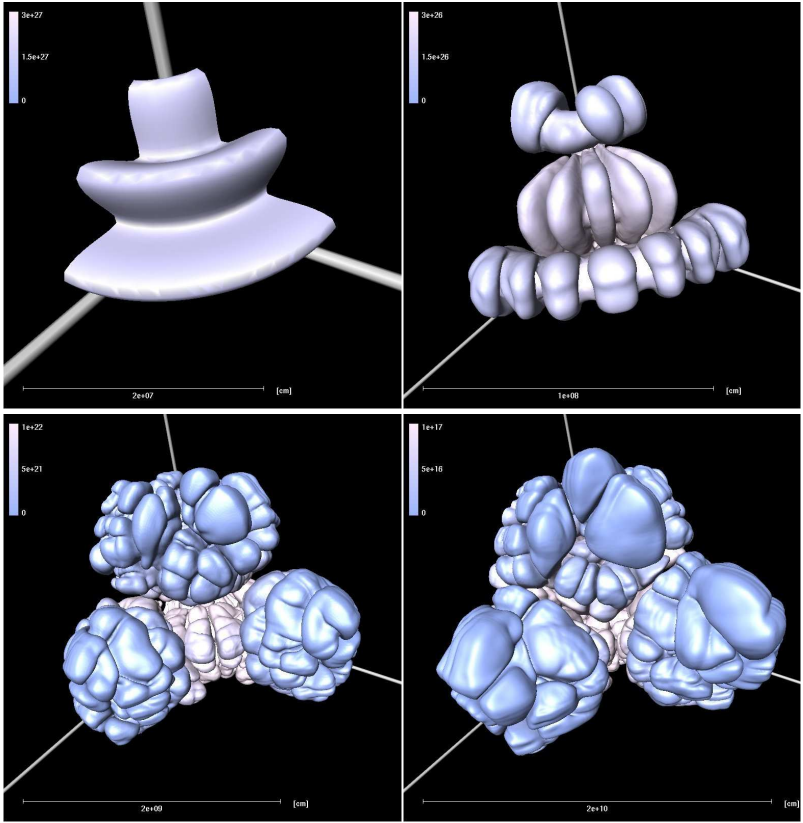}
}
\caption{Snapshots from a single-octant SN~Ia simulation performed on a
  uniform expanding computational grid. The isosurface corresponds to
  $G = 0$ and is
  color-coded with a measure of the turbulence strength derived from
  the subgrid-scale model (top left to bottom right:
  ignition flame, $0.5
  \, \mathrm{s}$, $2.0
  \, \mathrm{s}$, and $10.0
  \, \mathrm{s}$ after ignition).\label{fig:c3}}
\end{figure}

One requirement to reliably derive observables from simulations is that
the evolution of the models be followed to the stage of homologous
expansion. In this hydrodynamically relaxed situation, the velocity of
the ejected material is proportional to its radius.
Obviously, a static computational grid that sufficiently resolves the
flame propagation does not allow to follow the
expansion over sufficiently long time scales. Therefore \citet{roepke2005c}
proposed to use a moving computational grid that co-expands with the
WD. With this implementation it is in principle possible to follow the
evolution for arbitrary times, but \citep{roepke2005c} showed that
simulating the first $10 \, \mathrm{s}$ after ignition is sufficient
to reach homology with reasonable accuracy. An example of such a
simulation is shown in Fig.~\ref{fig:c3}, where
the isosurface represents the zero level set of $G$ and is associated
with the flame front. This simulation was carried out on only one octant
of the WD assuming mirror symmetry to the other octants. The flame was
ignited centrally with a toroidal perturbation superposed to the
spherical shape. Of course, at times
after $\sim$$2 \, \mathrm{s}$ when burning terminates in the model,
the zero level set of $G$ looses physical meaning, but it still
indicates the approximate boundary between unburnt material and
ashes.

\begin{figure}[t!]
\centering
\includegraphics[width = \linewidth, clip]{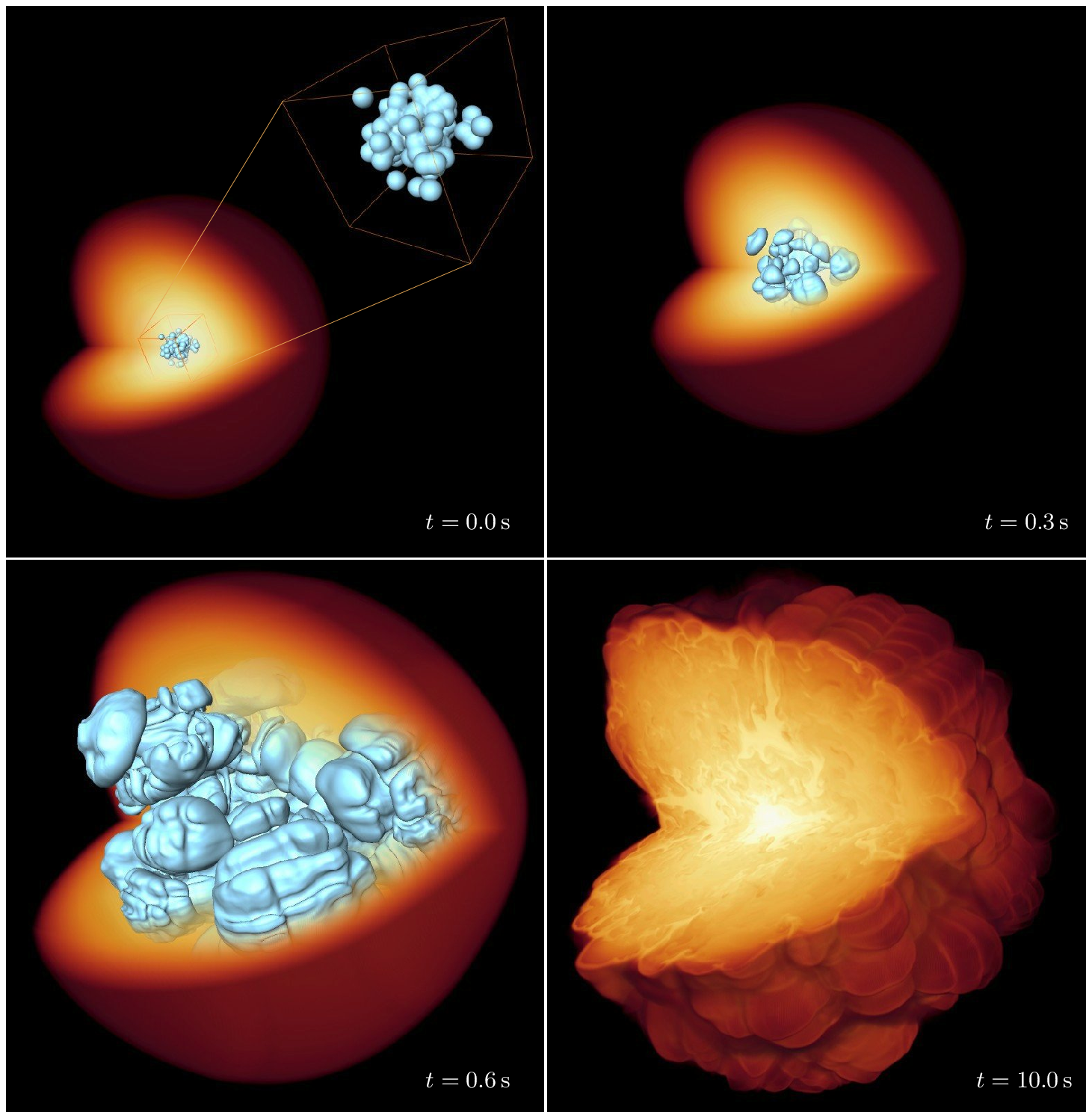}
\caption{Snapshots from a full-star SN~Ia simulation starting from
  a multi-spot ignition scenario. The logarithm of the density is
  volume rendered indicating the extend of the WD star and the
  isosurface corresponds to the thermonuclear flame. The last snapshot
  corresponds to the end of the simulation and is not on scale with
  the earlier snapshots.\label{fig:evo}}
\end{figure}

Due to the high computational expenses, most 3D
simulations (such as the one described above) comprise only one octant
of the WD. However, only full-star setups
allow to account for asymmetry effects.
On the basis of spectrapolarimetry observations of several SNe~Ia
\citep[e.g.][]{wang2003a}
these are expected to occur in at least some explosions.
\citet{roepke2005b} showed, that in the deflagration model such
asymmetries arise exclusively from irregularities in the flame ignition
conditions and not from large-scale instabilities and resulting
prefered modes in the flow patterns. Therefore, fixing a symmetric initial
flame shape and studying the influence of other physical parameters on
single-octant explosion models is a valid approach (see Sect.~\ref{sect:div} below).

To illustrate the typical flame evolution in deflagration SN~Ia models,
the full-star model presented by
\citet{roepke2005b} shall be described here. It started
out from an asymmetric initial flame configuration which was set up by
randomly distributing spherical flame kernels around the center of the
WD. This resulted in a foamy structure
slightly misaligned with the center of the WD.

Starting from this
initial flame configuration (shown in Fig.~\ref{fig:evo}), the
evolution of the flame front in the explosion process is illustrated
by snapshots of the $G=0$ isosurface at $t = 0.3 \, \mathrm{s}$ and $t =
0.6 \, \mathrm{s}$ in
Fig.~\ref{fig:evo}.
The development of
the flame shape from ignition to $t = 0.3 \, \mathrm{s}$ is
characterized by the formation of the well-known ``mushroom-like'' structures resulting
from buoyancy. This is especially well visible for the bubbles that
were detached from the bulk of the initial flame. But also the
perturbed parts of the contiguous flame closer to the center develop
nonlinear Rayleigh-Taylor like features.
During the following flame evolution, inner structures of smaller
scales catch up with the outer ``mushrooms'' and the initially separated structures
merge forming a more closed configuration (see snapshot at $t = 0.6 \,
\mathrm{s}$ of Fig.~\ref{fig:evo}). This is a result of the
large-scale flame advection in the turbulent flow and the expansion of
the ashes.
Up to this stage the flame was strongly anisotropic.
However, in the later evolution a preferentially lateral growth of
bubbles filled with ashes smoothes out parts of the anisotropies.
The flame develops a more spherical shape and only a slight anisotropy is
retained. After about $2 \, \mathrm{s}$ self-propagation
of the flame due to burning has terminated in the model.
The subsequent evolution is characterized by the approach to
homologous expansion. The resulting density structure at the end of
the simulation is shown in the $t=10\, \mathrm{s}$ snapshot of Fig.~\ref{fig:evo}.
The flame evolution agrees with the expectations
outlined in Sect.~\ref{sect:flame_explosion}.

Since the flame ignition process is not modeled in the explosion
simulations, there exists considerable freedom in choosing the number and
distribution of ignition kernels (cf.\
Sect.~\ref{sect:progenitor_ignition}). The results from models starting
with a central ignition (cf.\ Fig.~\ref{fig:c3}) indicate
that such a setup leads only to weak explosions. Recently, the
capabilities of multi-spot ignition models have been analyzed in a
systematic way. This was possible due to a modification of the moving
grid implementation to two nested grids. An inner, fine-resolved grid
follows the flame propagation while an outer coarse grid tracks the WD
expansion \citep{roepke2005e}. In this way, it is possible to
accumulate a large fraction of the available computational cells in the
inner regions thereby resolving detailed multi-ignition
configurations.
Confining the ignition volume around the
center of the WD, it may be expected that there exists a
certain number of ignition spots which maximizes the burning and
energy release. A sparse ignition would decrease the flame surface and
thus the burning rate, while a too dense distribution of flame kernels
will lead to a rapid merging of flame parts again decreasing the
surface area. This was confirmed by the simulations, which however did
not sharply single out an optimal number of ignition points but
revealed a rather robust behavior over a wide range. Extreme cases,
however, reproduced the anticipated dependence.

\subsection{Comparison with observations}
\label{sect:compare}

Due to recent progress in
deriving observables from multi-dimensional deflagration simulations, a direct
comparison with details of observations of nearby
SNe~Ia has come into reach.
Since these contain no other parameters than the initial
conditions, the question arises if the outcome of simulations as the
ones described above meets the observational constraints. Such
constraints result from the global characteristics derived from
observations, observed light curves, and spectra taken from nearby SNe
Ia.

The global characteristics derived from SN~Ia observations state that
a valid explosion model should release around $10^{51}\,\mathrm{erg}$
of energy and produce $\sim$$0.4\ldots 0.7\, M_\odot$ of $^{56}$Ni in
the nuclear burning \citep{contardo2000a,stritzinger2005a}. However, there exists a large diversity in the
observations ranging from the class of sub-luminous SNe~Ia (like
SN~1991bg with probably $\sim$$0.1 \, M_\odot$ of $^{56}$Ni) to
super-luminous events (e.g.\ SN~1991T with a $^{56}$Ni mass close to $1 \, M_\odot$). Deflagration
models started with a multi-spot ignition setup typically possess
$6\ldots 8 \times 10^{50}\,\mathrm{erg}$ of asymptotic kinetic energy
of the ejecta. They produce up to $\sim 0.4 \, M_\odot$ of
$^{56}$Ni. Thus they fall into the range of observational
expectations, but in the current stage do not account for the more
energetic SNe~Ia. One reason for the low energetics may be that in the
presented implementation nuclear burning is assumed to cease when the
fuel density drops below $10^7 \, \mathrm{g} \, \mathrm{cm}^{-3}$,
because the flame is then expected to enter the
distributed burning regime. An approach to handling this stage was
given by \citet{roepke2005a}.\\[-6mm]

\subsubsection{Lightcurves}

\begin{figure}[t]
\centerline{
\includegraphics[width = 0.7 \linewidth]
  {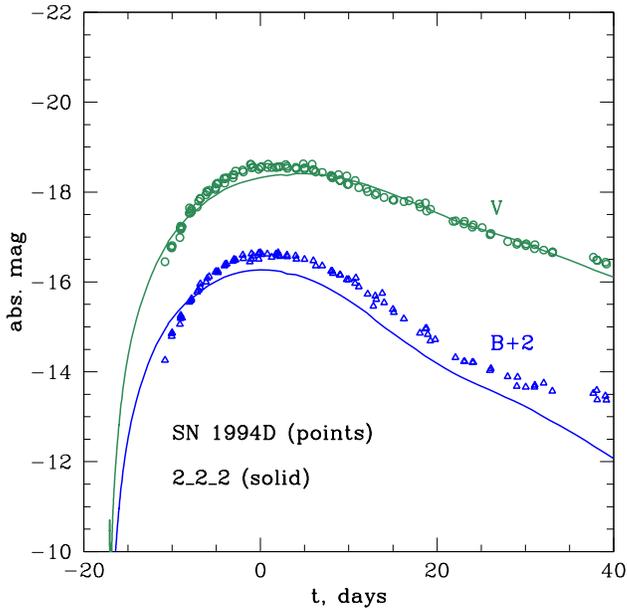}
}
\caption{Synthetic light curves derived from model 2\_2\_2 of
  \citep{roepke2005d} (solid curves) compared
  with observed light curves from SN~1994D. \label{fig:lc}}
\end{figure}
\vspace*{-1mm}
A further requirement is that synthetic lightcurves agree with
observed ones. These are sensitive to the energy release, the
$^{56}$Ni production, as well as to the distribution of elements in
the ejecta. In Fig.~\ref{fig:lc} synthetic light curves
derived from the \emph{2\_2\_2} simulation \citep{roepke2005d} are
compared with
observations of SN~1994D\footnote{The distance  to
  SN~1994D is still controversial and here the value of 30.4 is
  adopted from \citet{drenkhahn1999a}} (\citealp{blinnikov2005a}, see
also \citealp{sorokina2003a}).
The multi-band light curve of this model was
calculated using the \textsc{Stella} code of \citet{blinnikov1998a} and
\citet{blinnikov2000a}.

The model produced $0.3 \, M_\odot$ of $^{56}$Ni, although observations
require somewhat higher $^{56}$Ni mass ($\sim 0.4 \, M_\odot$) for the assumed
distance. Nonetheless, there is
generally very good agreement in the  B and V bands near
peak luminosities and in decline rate 20 days after the peak
which is most important for cosmological applications of type Ia supernovae.

\subsubsection{Spectra}

A much harder test for the models is posed by the comparison of
synthetic with observed spectra since these depend on details in the
composition of the ejected material.

\begin{figure}[t]
\centerline{
\includegraphics[width = 0.85 \linewidth, clip]
  {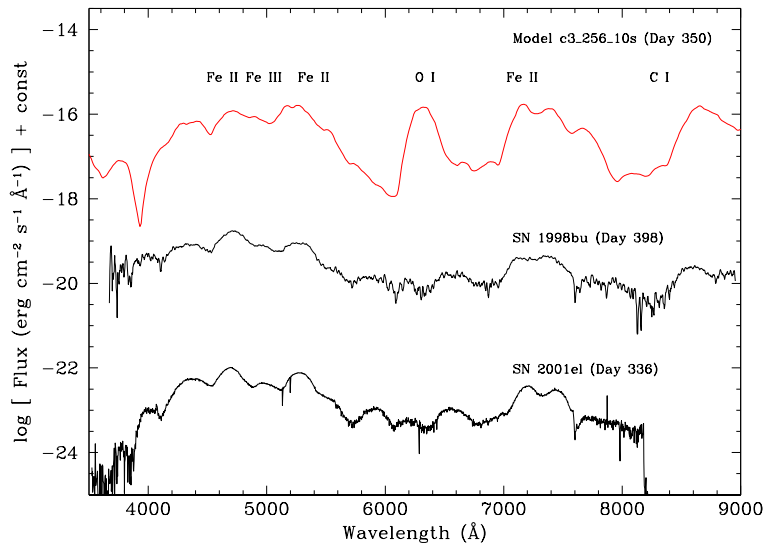}
}
\caption{Synthetic nebular spectrum compared with observations
  \citep[from][]{kozma2005a}.\label{fig:ns}}
\end{figure}

Nebular spectra provide a means of studying the central parts of the
ejecta, since they are taken at epochs where these have become
transparent due to expansion. Thus they explore the ``heart of the
explosion'' and are a valuable tool to study details of the physical
processes involved in the explosion stage. Unfortunately, only one
single synthetic late time spectrum is available from deflagration
SN~Ia models \citep{kozma2005a}. It was derived from the simulation
shown in Fig.~\ref{fig:c3}. The artificial and simple initial flame
shape chosen here
gives reason to not expect a good agreement between model and
observation. However, as shown in Fig.~\ref{fig:ns}, the broad iron
features of the observed spectra are qualitatively reproduced. An
inconsistency of the model with the observed nebular spectra is the
appearance of a pronounced oxygen line at $6300\,\mathrm{\AA}$. Both
features of the synthetic
spectrum share a common origin. The broad iron lines are caused by NSE
material that is transported in the uprising plumes of ashes and thus
distributed in velocity space. At the same time strong downdrafts
carry unburnt material towards the center of the WD.

The disagreement may in part be attributed to the simplicity of the
explosion model. Its highly symmetric initial flame shape with large
imprinted perturbations favors a pronounced evolution of
large-scale Rayleigh-Taylor features. This problem, however, is not
necessarily generic to all SN~Ia deflagration models. Multi-spot
ignitions may lead to a more complete burning in the central parts of
the WD. This was demonstrated by \citet{roepke2005e}. For an optimal
number of ignition bubbles (and a rather wide range around this
number), the central parts of the ejecta become dominated by iron
group elements. Whether the achievable
suppression of unburnt material at low velocities is sufficient to be
consistent with observations needs further study.

A powerful diagnostic tool to compare SN~Ia models with observations
is provided by the abundance tomography presented by
\citet{stehle2005a}. It makes
use of spectra taken from SN~2002bo with an extraordinary good time
coverage. Fitting this sequence of data with synthetic spectra
unveils the composition of the ejecta in velocity space slice by
slice, since the photosphere moves gradually inwards with the expansion
of the remnant. This abundance tomography of the ejecta can be
compared with results of 3D models, when averaged
over the angles. Qualitatively, the mixed composition of the ejecta
found by \citet{stehle2005a} is reproduced by deflagration SN~Ia
models in a natural way since these predict a distribution of burnt
material with the rising bubbles. A problem was, however,
that older predicted large unburnt material fractions in the central
parts of the ejecta in disagreement with the results of
\citet{stehle2005a}. A recent high-resolved simulation cures
this problem by clearly reproducing the iron-group dominance in the
low-velocity ejecta \citep{roepke2005f}.

\subsection{Diversity and correlations}

\label{sect:div}

\begin{table}
  \begin{center}
    \caption{Variation of initial parameters in SN~Ia explosion
      models.}\vspace{1em}
    \renewcommand{\arraystretch}{1.25}
    \begin{tabular}[h]{lrrr}
      \hline
      Parameter & \multicolumn{1}{p{0.21 \linewidth}}{range of
      variation} &
      \multicolumn{1}{p{0.17 \linewidth}}{effect on $^{56}$Ni
      production} & \multicolumn{1}{p{0.17 \linewidth}}{effect on
      total energy} \\
      \hline
      X($^{12}$C)  & [0.30,0.62]  & $\le$$2\%$ & $\sim$14\% \\
      $\rho_\mathrm{c}$ [$10^9 \, \mathrm{g}/\mathrm{cm}^3$]  &
      [1.0,2.6] & $\sim$6\% & $\sim$17\% \\
      $Z$ [$Z_\odot$] & [0.5,3.0] & $\sim$20\% & none \\
      \hline \\
      \end{tabular}
    \label{tab:res}
  \end{center}
\end{table}

The recent developments in the deflagration SN~Ia explosion modeling
outlined in the previous section seem to indicate that such a model is
capable of reproducing the main features of observed objects but they do
not rule out the alternative of a delayed detonation.

Here, the question of how multi-dimensional SN~Ia models can be applied
to tackle questions from the cosmological applications of these
objects shall be illustrated on a set of simple deflagration models.
Of particular interest is how such models behave under variation of
physical parameters. Do they reproduce the observed
diversity of SNe~Ia? Are correlations between observables evident in
the results?

Unfortunately, 3D deflagration models of SNe~Ia as
described above are computationally expensive. To moderate these
expenses, simplified setups may be used to study effects of physical
parameters on the explosion models. Such an approach was recently
taken by \citet{roepke2005d} and resulted in the first systematic study of
progenitor parameters in 3D models. The basis of this
study was a single-octant setup with moderate (yet numerically
converged) resolution. However, the lack in resolution did not allow a
reasonable multi-spot ignition scenario and thus only weak explosions
can be expected. It was therefore not possible to set the absolute
scale of effects in this approach, but trends can clearly be
identified.

The parameters chosen for the study were the WD's
carbon-to-oxygen ratio, its central density at ignition and its
$^{22}$Ne mass fraction resulting from the metallicity of the
progenitor. All parameters were varied independently to study the
individual effects on the explosion process. In a realistic scenario,
however, these parameters are interrelated by the evolution of the
progenitor binary system. The results of this survey are given in
Tab.~\ref{tab:res}.

\begin{figure}[t]
\centerline{
\includegraphics[width = 0.98 \linewidth]
  {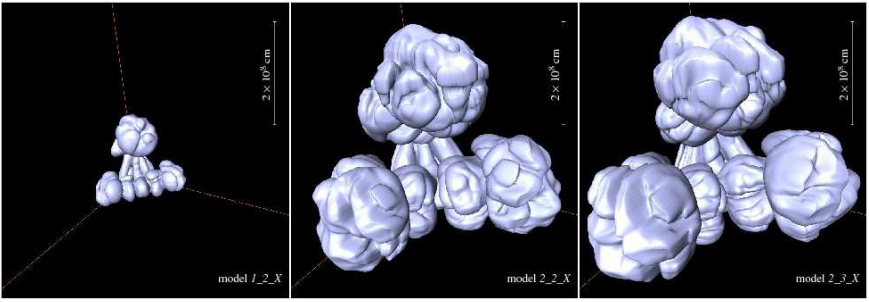}
}
\caption{Snapshots of the flame front evolution at $t = 1.0 \,
  \mathrm{s}$ after ignition from models with $\rho_c = 1.0 \times
  10^9 \, \mathrm{g} \, \mathrm{cm}^{-3}$ and $X(^{12}\mathrm{C}) =
  0.42$ (left);
  $\rho_c = 2.6 \times
  10^9 \, \mathrm{g} \, \mathrm{cm}^{-3}$ and $X(^{12}\mathrm{C}) =
  0.42$ (middle);
  $\rho_c = 2.6 \times
  10^9 \, \mathrm{g} \, \mathrm{cm}^{-3}$ and $X(^{12}\mathrm{C}) =
  0.62$ (right) \label{fig:grid_ag}}
\end{figure}

A variation of the carbon-to-oxygen ratio affects the energy production
in the burning due to the differences in the binding energies of these
two nuclei. Counterintuitively, this results in no significant change
in the $^{56}$Ni production. This finding is consistent with the flame
very similar flame evolution in simulations with different carbon mass
fraction of the WD (cf.\ Fig.~\ref{fig:grid_ag}). \citet{roepke2004c} explained this effect by the fact
that the potentially higher energy release in carbon-rich models is
buffered by a higher $\alpha$-particle fraction in the NSE material and only
released when burning is already incomplete.

Models with lower central densities show a delayed flame
evolution (see Fig.~\ref{fig:grid_ag}) and consequently a lower and delayed energy production. This
is due to the fact that the flame experiences a lower gravitational
acceleration in these models resulting in decreased turbulence
generation. Therefore less $^{56}$Ni is produced in these models. This
effect is even more pronounced due to the fact that in low-density WDs less
material is present at sufficient densities to be potentially burnt to
NSE. A counteractive effect is expected at higher densities. Here,
electron captures will become important favoring neutron-rich isotopes
over $^{56}$Ni in the NSE. The dynamical effects of electron captures are,
however, not yet implemented in the explosion model and therefore the
survey of \citet{roepke2005d} is restricted to relatively low central densities.

\begin{figure}[t!]
\centerline{
\includegraphics[width = 0.495 \linewidth]
  {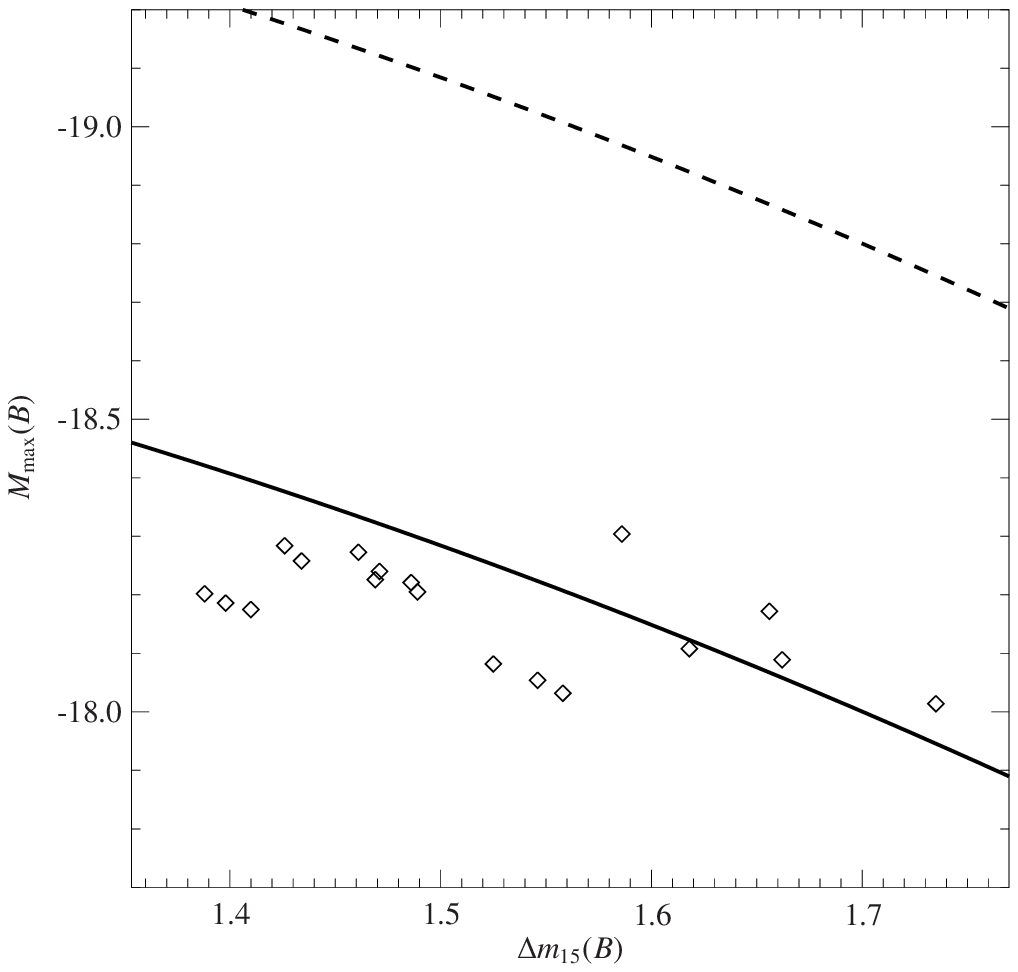}
\includegraphics[width = 0.495 \linewidth]
  {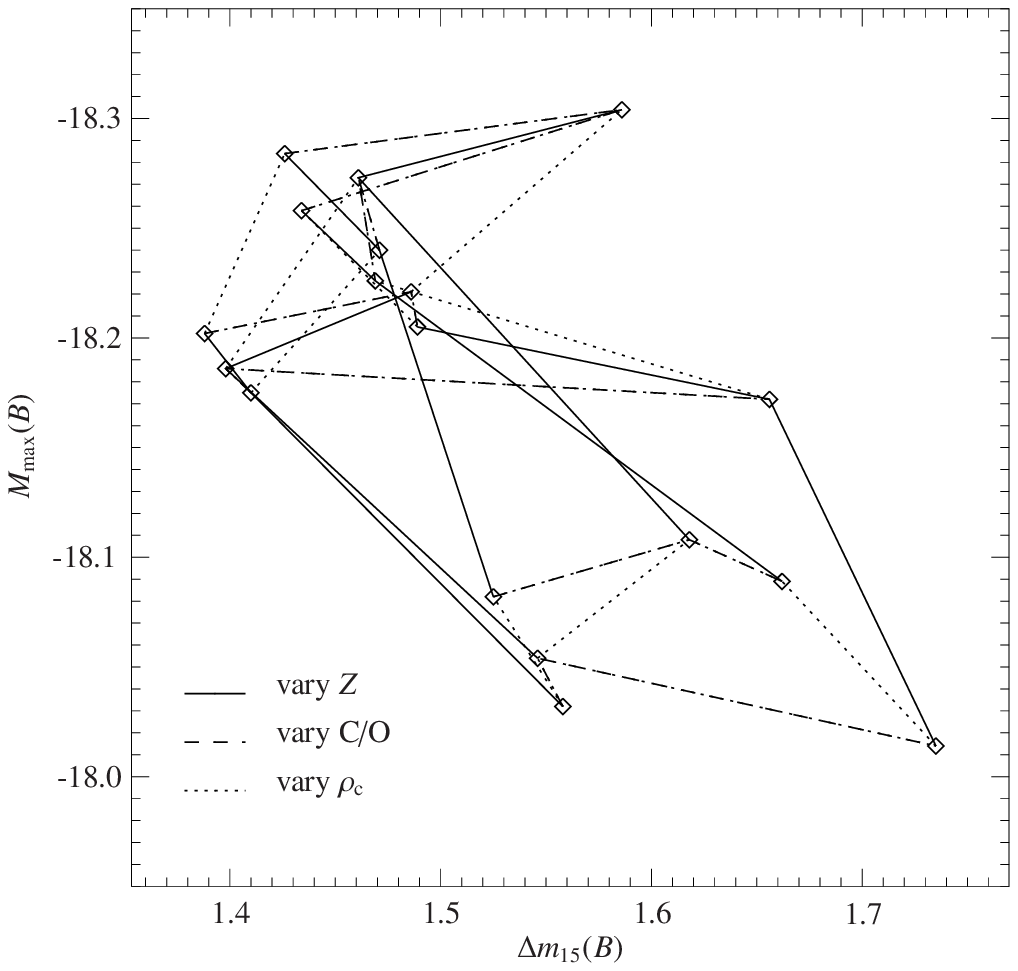}
}
\caption{Peak luminosity vs. decline rate of the light curve in the B
  band (diamonds correspond to SN~Ia explosion models). Compared with
  original relation by \citet{phillips1999a}
  (dashed curve) and shifted relation (solid curve) in the left
  panel.\label{fig:ph}}
\end{figure}

A higher metallicity of the main sequence progenitor star results in an
increased $^{22}$Ne
mass fraction in the WD. This is an isotope with neutron excess and
therefore again favors the production of neutron-rich species over
$^{56}$Ni in the NSE. The results of \citet{roepke2005d} confirm the analytic prediction by
\citet{timmes2003a} and agree with \citet{travaglio2005a}. The
metallicity parameter, however, has no effect on the explosion
dynamics and the energy production in the models.

To determine the effects of these variations on observables, synthetic
light curves were derived from all models (an example is shown in
Fig.~\ref{fig:lc}). From these, the peak
luminosities and decline rates (in magnitudes 15 days after maximum;
$\Delta m_{15}$), were determined. The pioneering work by
\citet{phillips1993a}
established the relation between the peak luminosity and $\Delta
m_{15}$ as one of the primary tools to calibrate cosmological SN~Ia
distance measurements. The so-called \emph{Phillips relation}
quantifies the decrease of $\Delta m_{15}$ for brighter SNe~Ia.

The results from the presented deflagration models are compared with
the relation given by
\citet{phillips1999a} in Fig.~\ref{fig:ph}. Obviously, the absolute magnitude of the
\citet{phillips1999a} relation is not met by the set of models (cf.\ the upper panel of
 Fig.~\ref{fig:ph}). Moreover, the range
of scatter in $\Delta m_{15}$ is much narrower than that of the set of
observations used by \citet{phillips1999a}. The simulations exhibit a
large scatter but are consistent with the slope of the
\citet{phillips1999a} relation. It is, however, obvious, that a better
agreement
cannot be expected given the fact, that the parameters in the survey
were chosen independently. A consistent stellar evolution would pick a
sub-sample of the set of models possibly narrowing the range of scatter.

Nonetheless, with the set of models, the question can be answered,
which parameter dominates the slope in the direction of dimmer events
for faster decline rates. The varied parameters are coded by different
line styles in the right panel of Fig.~\ref{fig:ph}. Clearly, the
progenitor's metallicity can be identified as this
parameter. Variations in the central density and the carbon mass
fraction of the WD superpose a scatter on the dominant relation.

\section{Conclusions}
\label{sect:concl}

The rapid development of multi-dimensional SN~Ia models over the past
years led to an better understanding of the mechanism of these
events. In particular, it has been shown, that 3D SN~Ia
simulations on the basis of the deflagration model are capable of
reproducing main features of observed SNe~Ia. However, currently it
seems that they can only account for the weaker objects. A better
determination of the initial conditions, i.e.\ the progenitor
evolution and flame ignition, may improve the situation. It may,
however, also turn out that the current model is incomplete. A delayed
detonation was suggested by several authors to improve the agreement
with observations, lacking however a convincing mechanism for a
deflagration-to-detonation.

On the exemplary case of a particular numerical implementation of the
deflagration scenario it was shown how such a model can be validated
against observations and applied to determine the origin of the
diversity of SNe~Ia as a first step to theoretically assess the
empirical relations utilized in SN~Ia cosmology.

\end{document}